\documentstyle[pre,tighten,preprint,aps]{revtex}
\input{psfig.sty}
\newcommand{\eps}{\varepsilon}

\newcommand{\dfrac}[2]{\frac{\normalsize #1}{\normalsize #2}}
\newcommand{\mod}{\mbox{mod}}

\draft
\title{Synchronization and
	directed percolation in  
	coupled map lattices
}

\author{Franco Bagnoli${}^{(1,4)}$\cite{bagnoli}}
\author{Lucia Baroni$^{(2,4)}$}
\author{Paolo Palmerini$^{(3)}$}
\address{1) Dipartimento di Matematica Applicata,
		Universit\`a di Firenze,  Via S. Marta 3, I-50139 Firenze,
		Italy}
\address{2)  Dipartimento di Fisica, Universit\`a di Firenze,
		Largo E. Fermi 2, I-50125 Firenze, Italy.}
\address{3) CNUCE-CNR, via S. Maria 36, I-56100 Pisa, Italy}
\address{4) INFN and INFM, Sez. di Firenze.}

\date{\today}
\begin{document}
\maketitle

\begin{abstract}                
	We study a synchronization mechanism,  based on one-way coupling
	of all-or-nothing type, applied to coupled map lattices with
	several different local rules. By analyzing the metric and the
	topological distance between the two systems, we found two
	different regimes: a strong chaos phase in which the transition
	has a directed percolation character and a weak chaos phase in
	which the synchronization transition occurs abruptly.  We are
	able to derive some analytical approximations for the location of
	the transition point and the critical properties of the system.
	We propose to use the characteristics of this transition as
	indicators of the spatial propagation of chaoticity.
\end{abstract}


\section{Introduction}
Recently, the synchronization of chaotic
systems has received considerable attention~\cite{PecoraCarroll,%
Gupteetal,Maritan,DingOtt,RosenblumPikowsky,%
Kocarev,MorgulFeki}. 
Among the many papers on this subject, some of them
concern the behavior of spatially extended chaotic 
systems~\cite{Heagyetal,Kocarevetal,Parmananda}. 

In this paper we study the synchronization properties of two coupled map
lattices~\cite{Kaneko}, when the coupling between them  is completely
asymmetric,  and the synchronization mechanism is of all-or-none kind: denoting
one system as the master, and the other as the slave one,   either an
individual map of the slave system is completely synchronized to the
corresponding map of the master system, or it is left free. We study the
annealed version of this kind of coupling, by choosing at random a fixed 
fraction of  sites to be synchronized at each time step, which constitutes the
control parameter of the synchronization transition. 

An alternative point of view origins from the problem of the characterization
of chaos in high dimensional  dynamical systems.  We assume a rather unusual
point of view: instead  of considering the rate of divergence of the distance
between the  trajectory of a reference system (the master) and a  perturbed one
(the slave), procedure that leads to the definition of the maximum Lyapunov
exponent, we measure the ``efforts'' needed to make the slave system coalesce
with the master. 

Let us illustrate in detail this approach. The standard technique for studying
the chaotic properties of such a system is that of measuring the response of
the system to a perturbation. The conceptual experiment is the following: (1) 
at a certain time make a copy of the master system and slightly vary its state;
(2)  let the system evolve for a small interval of  time and (3) measure
the distance between the master and the slave; (4) renormalize the distance so
that it is always small. The logarithmic average of the growth rate of the
distance gives the maximum Lyapunov exponent.  For a small distance,  the
leading contribution in the diverging rate comes from the maximum eigenvalue of
the product of the  Jacobian of the evolution function computed  along the
trajectory. By considering more than just one vector in the tangent space, and
keeping them orthogonal, one can obtain the whole spectrum of Lyapunov
exponents~\cite{Benettin}.

The Lyapunov spectrum, however, does not describe accurately the process of
spatial propagation  of chaos. We shall illustrate this point  by considering 
diffusively coupled logistic map lattices at Ulam point.  
 By analyzing the behavior of the Lyapunov  spectrum, one
sees that for a small coupling the spectrum is almost constant and positive.
When increasing the coupling between maps, both the positive part of the
Lyapunov spectrum and the maximum exponent decrease. 

From a different point of view, the chaoticity (and the Lyapunov exponent) of a
single map can be defined by means of the ``efforts'' needed to synchronize
the  slave with the master. Let us consider the following ``toy'' system
composed of two simple maps
\begin{equation}
	\label{singlemap}
	\left\{\begin{array}{cl}
		x' &= f(x);\\
		y' &= (1-p) f(y) + p f(x).
	\end{array}\right.
\end{equation}
where $x \equiv x(t)$  is the master system and $y \equiv y(t)$ is the
slave~\cite{Pikowski}; the prime denotes the value of the map at time $t+1$.  
The master system $x$ evolves freely, while the slave $y$ is  subjected to two
opposed contributions: it tends to separate from the master if the  map $f$ is
chaotic, but it is pushed towards $x$ by the parameter $p$, which represents
the ``strength''.

For a small difference  $z(t) = x(t)-y(t)$ one has, 
\begin{equation}
	z' = (1-p) f'(x) z \label{singlemaptangent}
\end{equation}
i.e.\ the synchronization ($z(\infty)=0$) occurs  for $p \ge
p_c=1-\exp(-\lambda)$, where $\lambda$ is the Lyapunov exponent of the
(unperturbed) trajectory $x(t)$ of the  master system. Thus for simple maps the
synchronization threshold is related to the chaoticity of the system. We would
like to extend this concept to spatially extended system, with a diffusive
character. 

The basic idea of our approach  is to consider Eq.~(\ref{singlemap}) as a mean
field description of a stochastic process which, with probability $p$, sets
each individual map of the slave system  to the value of the corresponding map
of the master one (``pinching''). We assume that the  synchronized state is
absorbing, i.e. once a patch of the slave system is in the same state of the
master one, desynchronization can occour only at borders, without bubbling. The
completely synchronized state, even if unstable, cannot be exited. A discussion
about the robustness of this synchronization mechansm is deferred to the last
Section.

This method can be applied also to more exotic dynamical systems, such as 
cellular automata~\cite{BagnoliRechtman}. For such  systems the usual chaotic
indicators cannot be easily computed, but they exhibit spatial propagation of
disorder. 

It is clear that for $p=1$ the system synchronizes in just one time step. 
However, as we shall show in details in Section~3, a critical value  $p=p_c<1$ 
does exist, above which the system synchronizes regardless of its chaoticity.
This threshold is related to a directed percolation phase transition.

On the other hand, some systems can synchronize for $p=p^*<p_c$, because the
propagation rate of the differences between the master and the slave is reduced
by the inactivation of the degrees of freedom due to pinching.  As a trivial
illustration of this process, a lattice of \emph{uncoupled} maps synchronizes
in the long time limit for all strengths $p>p^*=0$. This synchronization
threshold is an  indicator of the spatial propagation of chaoticity in the
system. Other informations  come from the dynamics of the transition.

The sketch of this paper is the following: in the next Section we describe
precisely the model we use and introduce the observables, in Section~3 the
synchronization transition is analyzed, and the connections with the directed
percolation problem discussed. In Section~4 we present the phase diagrams for
the synchronization transition for several well-known maps, and finally in the
last Section we discuss our main conclusions, including possible extensions 
and applications of the synchronization mechanism, and the relationships
with other models found in literature.   


\section{The stochastic synchronization mechanism} 
\label{sec:themodel}

The state of the master system at a given time $t$ is denoted as  $X^t \in
[0,1]^N$, and a component of $X$ (a single map) is  indicated as $X|_i \equiv
x_i$, $i=1,\dots,N$.  The dynamics is defined as 
\begin{equation}
	x^{t+1}_i =  g_\eps(x_{i-1}^t, x_i^t, x_{i+1}^t) \equiv
		(1-2\eps) f( x_i^t) +
		\eps\big(f(x_{i-1}^t) + f(x_{i-1}^t)\big),\label{xevol}
\end{equation}
with periodic boundary conditions and $0 \le \eps \le 1/2$. It can be
considered as the discretization (in time and space) of a reaction-diffusion
system, the $\eps$-coupling coming from the Laplacian operator.
The dynamics of the slave system $Y$ is given by 
\begin{equation}
		y^{t+1}_i =  (1-r_i^{t}(p)) g_\eps (y_{i-1}^t, y_i^t, y_{i+1}^t) +
			 r_i^{t}(p) g_\eps(x_{i-1}^t, x_i^t, x_{i+1}^t), \label {yevol}
\end{equation}
where $r_i^t(p)$ is a random variable  which assumes the value one with
probability $p$, and zero otherwise. In other words,  at each time step a
fraction $p$ of maps in the slave system  is set  to the same value of the
corresponding map in the master system. 

In vectorial notation one can write for the master system
\[
	X^{t+1} = G_{\eps} (X^t) = (I+\eps \Delta) F(X^t),
\]
where $\Delta$ is the discrete Laplacian 
\[
	\Delta X|_{i j} = \big(\delta_{i j-1}+\delta_{i j+1}
	-2\delta_{i j}\big)x_j
\]
and $F(X)$ is a diagonal operator $F(X)|_{ij} = f(x_j)\delta_{i j}$. For the
slave system one has
\[
	Y^{t+1} = \overline{S}^t(p) G_\eps (Y^t) + S^t(p) G_\eps(X^t)
\]
where $S(p)$ is a random diagonal matrix having a fraction $p$ of diagonal
elements equal to one and all others equal to zero, $S(p)^t|_{ij} = r_j^t(p)
\delta_{ij}$,  and  $\overline{S}(p) = I-S(p)$.

We introduce also the difference system $Z=X-Y$, whose evolution rule
is 
\[
	Z^{t+1} = \overline{S}^t(p) \bigl(G_\eps(X^t) - G_\eps(X^t-Z^t)\bigr)
\]

If the difference field is uniformly small, i.e.\ 
for $\max (|z_i^t|)\rightarrow 0$, one has  
\[
	Z^{t+1}   = \overline{S}^t(p) J_\eps(X^t) Z^t,
\]
where $J_\eps(X)$ is the Jacobian of the evolution function
\[
	J_\eps(X^t) = (I+\eps \Delta) \dfrac{\partial F(X^t)}{\partial X^t}
\]
and 
\[
	\left.\dfrac{\partial F(X)}{\partial X}\right|_{i j} = 
	\delta_{i j} \left. \dfrac{d f(x)}{dx}\right|_{x=x_j} .
\]

For all kinds of numerical computations  there is a limit to the precision
below which two numbers become indistinguishable. Since this limit depends on
the magnitude of the numbers, it is very hard to control its effects. In this
perspective, we introduce a threshold $\tau$ on the precision, by imposing that
if $|x_i-y_i|<\tau$ then $y_i=x_i$. In this way we can study the sensitivity of
the results on $\tau$, and eventually perform the limit $\tau\rightarrow 0$. We
have checked that our asymptotic results are independent of $\tau$, at least
for $\tau$ smaller than $10^{-6}$. In the following we shall neglect to
indicate the truncation operation for the ease of notation, except when
explicitly needed. Due to the precision threshold $\tau$, the difference field
$z_i^t$ is set to zero if $z_i^t<\tau$. 

An alternative way of computing the evolution of the  system, which will be
useful in the following, is given by the following procedure:
\begin{enumerate}
\item Consider three stacked two-dimensional lattices of size $N\times T$, 
	and label one direction as space $i$, $i\in [1,N]$ and the other one as
	time $t$, $t \in [0,T]$. This ensemble can be considered composed by three
	layers: one that contains the Boolean numbers $r_i^t$, and two containing
	the  real numbers $x_i^t$ and $y_i^t$.
\item Fill up the $r_i^t$ layer with zeros and ones so 
	that the probability of having $r_i^t=1$ is $p$; this layer will be named
	the quenched field.
\item Fill up the $x_i^0$ and $y_i^0$ rows at random;
	they will be the initial conditions for the $X$ and $Y$ lattice maps. 
\item Iterate the applications (\ref{xevol}) and (\ref{yevol}) to fill
	up the $X$ and $Y$ layers.
\end{enumerate}

In this way we define an out of equilibrium statistical system, with $p$ as a
control parameter. It is assumed the limit $N\rightarrow\infty$ and
$T\rightarrow\infty$ and the average over the quenched field.  The degree of
synchronization of the system at a given time $t$ can be measured by the
(metric) distance
\[
	\zeta^t=\dfrac{1}{N}  \sum_{i=0}^{N-1} |z_i^t|,
\]
whose asymptotic value for a given probability $p$ will be denoted as
$\zeta(p)$. 

We introduce also the field $h_i^t$ as
\[
	 h_i^t=\left\{\begin{array}{cl}
		  0 &\text{if $z_i^t =0$;}\\
	    1 &\text{otherwise,}
	 \end{array}\right.
\]
and the topological distance 
\[
	\rho^t=\dfrac{1}{N}  \sum_{i=1}^{N} h_i^t
\]
which measures the fraction of non-synchronized sites in the system. The
asymptotic value of the topological distance will be denoted as $\rho(p)$.

We shall study the synchronization transition  for  the following maps $f(x)$
of the unit interval $x\in [0,1]$:
\begin{enumerate}
\item The generalized Bernoulli shift $f(x)= [ax] \mod 1$ with a slope
	$a$ greater than one. The Lyapunov exponent of the single map is simply
	$\ln(a)$; this is also the value of the maximum Lyapunov exponent for  a
	diffusively  coupled lattice, regardless of $\eps$.   \item The quenched
	random map that assumes a different (random) value for each different $x$.
	It can be considered equivalent to the Bernoulli shift  in the limit
	$a\rightarrow\infty$, and thus the map with the highest degree of
	chaoticity. This map is everywhere non-differentiable.
\item The logistic map $f(x) = a x (1-x)$, which is chaotic for 
 	$3.57\dots < a \le 4$. 
\item The generalized tent map 
	\[
		f(x)=\left[a\left(\dfrac{1}{2}-\left|x-\dfrac{1}{2}\right|\right)
		\right] \mod 1
	\]
	with $a>1$.
\end {enumerate}

\section{The synchronization transition}
\label{sec:transition}
If the  maps are uncoupled ($\eps=0$), any value of $p$ greater than zero is
sufficient to synchronize the system in the long time limit, regardless of the
chaoticity of the single map.  For coupled systems, however, the chaoticity of
the map (temporal chaos) contributes  to the spatial chaoticity of the system. 

\begin{figure}
		\centerline{\psfig{figure=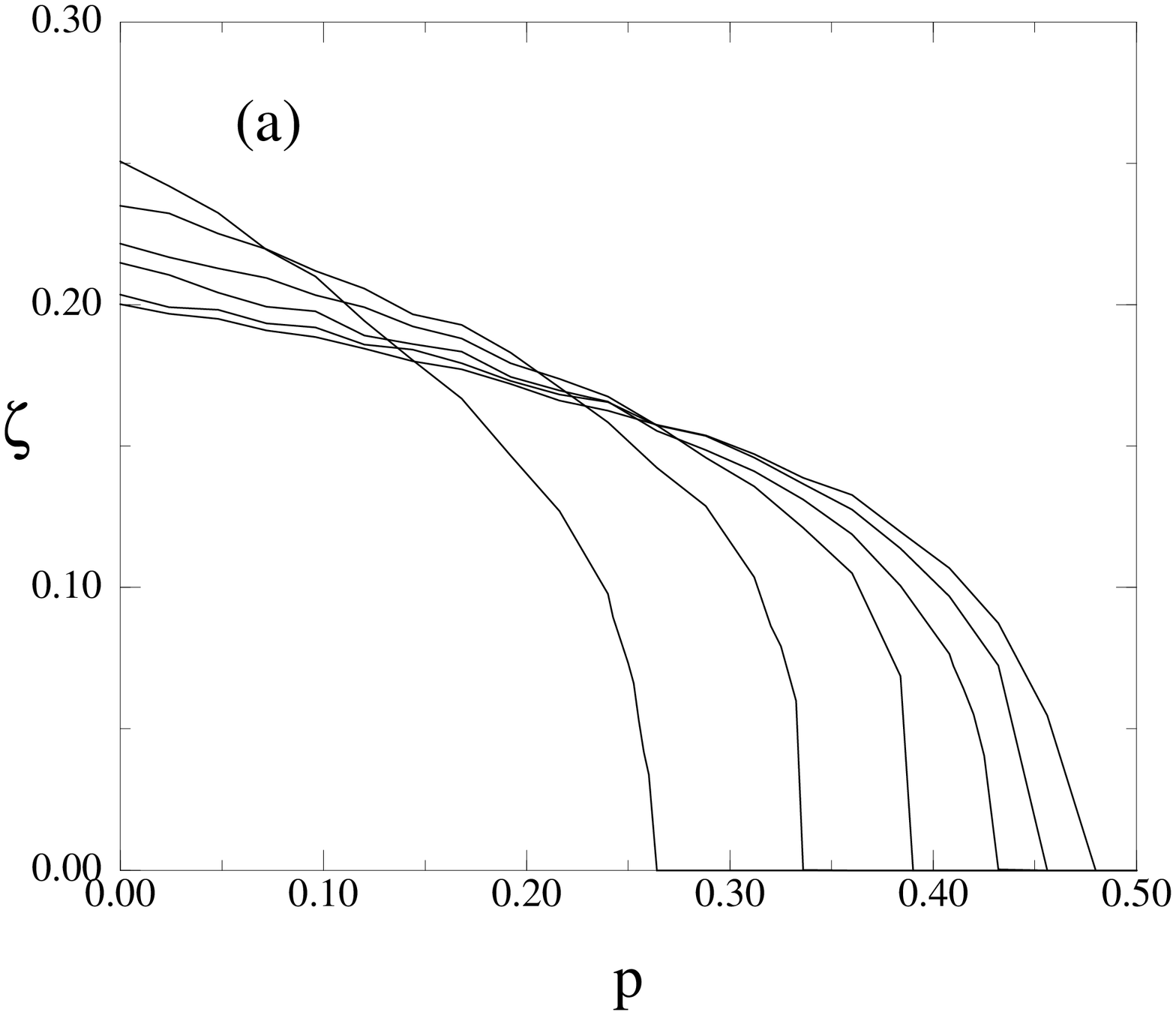,angle=270,width=11cm}} 
		
		\centerline{\psfig{figure=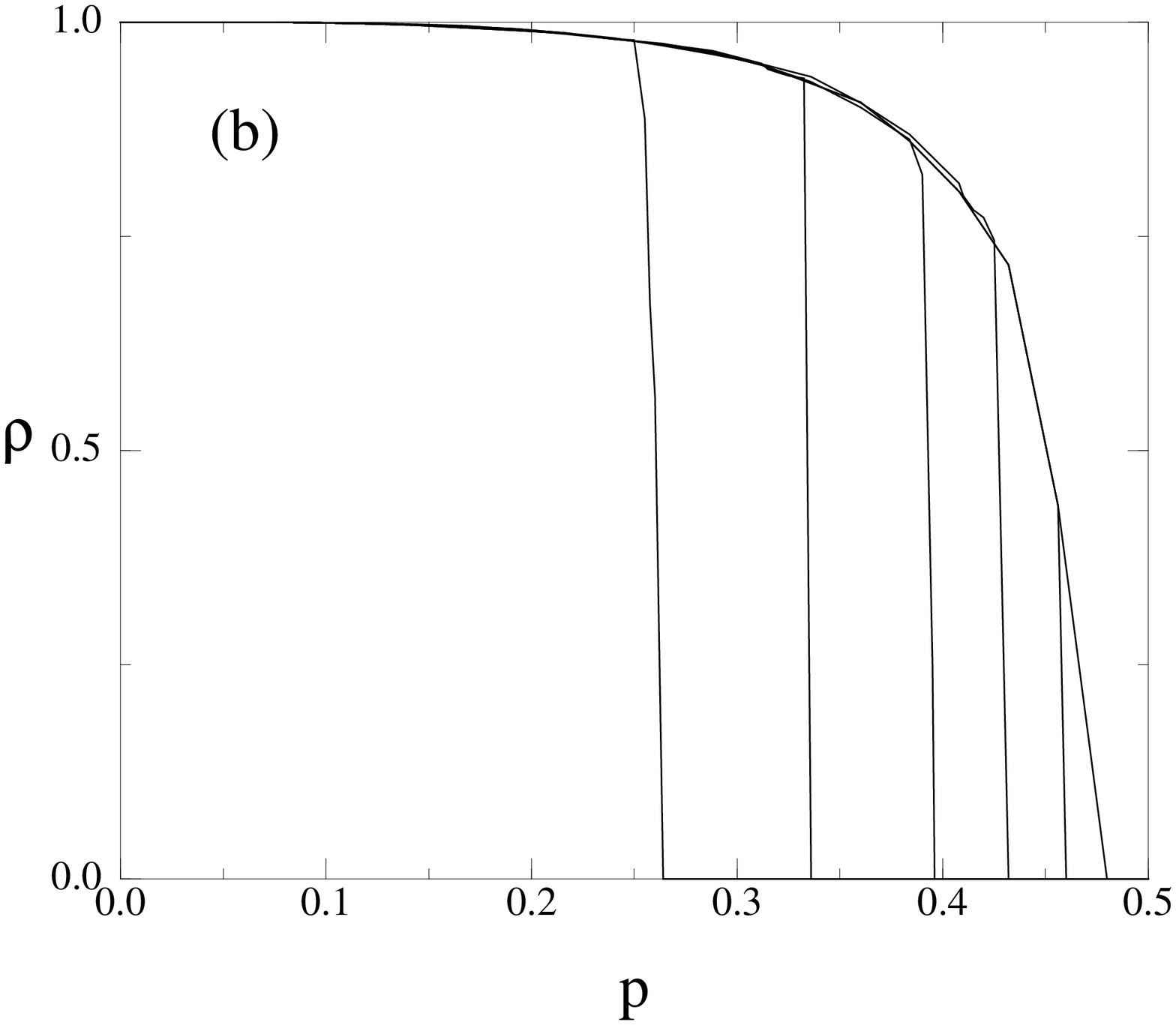,angle=270,width=11cm}} 
	
	\caption{\label{fig:rhozetaBernoulli}
		Metric distance (a) and topological distance (b) 
		for a chain of 32000 Bernoulli maps
		$\eps=1/3$ and $a=1.2, 1.4,\dots, 2.2$ from left to right.
		One run of 30000 time steps.
	}
\end{figure}

For illustration purposes, we show in Fig.~\ref{fig:rhozetaBernoulli} the 
behavior of the metric distance $\zeta(p)$ (left) and the topological distance
$\rho(p)$ (right) for Bernoulli maps with $\eps=1/3$ and different values of
$a$.  The topological distance exhibits a sharp transition for $a<a_c\simeq 2$,
and a smooth transition for $a\ge a_c$; all curves superimposes to a universal
curve  far from the transition point. The metric distance always exhibits a
smooth transition. This scenario is generic for all kinds of maps and
couplings, except that some maps (notably the logistic map) never exhibit the
smooth transition of the topological distance in the allowed range of values of
the parameter $a$.

For the random map it is quite easy to understand the origin of the universal 
curve: since even a small distance is amplified in one
time step to a random value, the difference $z_i^{t}$ is greater than 0 if
$z_j^{t-1}>0$ on some of the neighbors ($|i-j|<1$) at the previous time step,
and the quenched field $r_i^{t}$ is equal to 0. 

\begin{figure}
	\centerline{\psfig{figure=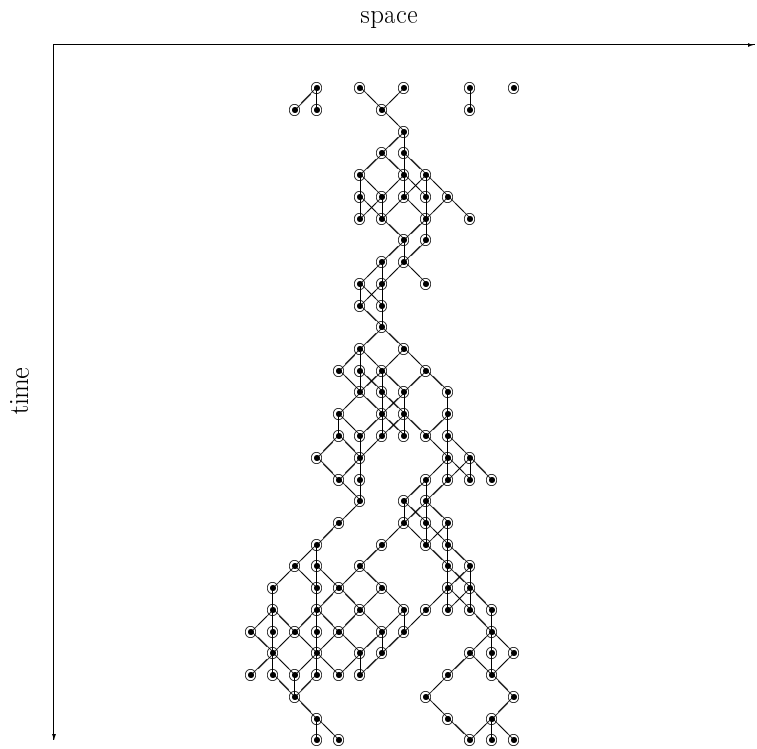,width=10cm}}
	\caption{An example of a wet cluster for $p=0.45$.}
	\label{fig:cluster}
\end{figure}

This defines a directed site percolation problem~\cite{DP} (formation of a
percolating cluster of ``wet'' sites along the time direction), with control
parameter $1-p$. A site of coordinate $(i,t)$ is said to be wet if $r_i^t(p)=0$
and it is connected at least to one neighboring wet site $r_j^{t-1}(p)=0$ 
($|i-j|\le 1$) site at time $t-1$. All sites in the first ($t=0$) row are
supposed to be connected to an external wet site. We shall denote the ensemble
of wet sites with the name \emph{wet cluster}; an instance of a wet cluster for
$p=0.45$ is shown in Fig.~\ref{fig:cluster}. Notice that the usual directed
percolation probability $p^{(\mbox{\scriptsize DP})}$ is given here by
$p^{(\mbox{\scriptsize DP})} =  1-p$. 

If $p^{(\mbox{\scriptsize DP})} < p^{(\mbox{\scriptsize DP})}_c$ the wet
cluster does not percolate in the time direction, and the system will
synchronize despite any chaoticity of the single map. The $r_i^t=0$ sites
disconnected from the percolation cluster do not influence the synchronization,
since the synchronized state is locally absorbing. Thus, for the random map and
$\eps>0$,  the invariant curve is simply the curve of the asymptotic density
$m(p)$ of wet sites for the directed percolation problem. In the vicinity of
the transition, $m(p)$ behaves as 
\[
	 m(p) \sim (p_c-p)^\beta,
\]
where $\beta\simeq 0.26(1)$ is the magnetic critical exponents of the $1+1$
dimensional directed percolation problem, and $p_c = 1-p_c^{(\mbox{\scriptsize
DP})} \simeq 0.460(2)$  for this lattice with connectivity
three~\cite{BagnoliBoccaraPalmerini}.

In the generic case, one has to study what happens on the wet cluster. If the
evolution of the system is expanding in  the difference space, and the wet
cluster percolates in the time direction, then the asymptotic topological
distance is greater than zero and vanishes  when the density of wet sites does.
The expansion rate in the difference space depends on the average number of
connections between wet sites in the cluster, an exception being the random
map, which expands to a random value regardless of the initial difference.

On the other hand, it may happen that for some value of $p$ (for a given
percolating wet cluster) the dynamics in the difference space is contracting,
so that the difference field $z_i^t$ will eventually become less than $\tau$
and thus set to zero. As an illustration, let us consider the Bernoulli shift
for $\eps=1/3$ and only one connection in average. The problem reduces to the
computation of the expansion rate in a one dimensional chain of asymmetrically
coupled maps, and the threshold for the synchronization is $a=3$.

This eventual contraction mechanism implies that the cluster of the sites for
which $h_i^t=1$ (the difference cluster) is generally strictly included into
the percolation cluster.

We classify the case in which the difference cluster always survives  when the
percolation cluster spans the lattice with the name  \emph{strong
spatiotemporal chaos}, otherwise  we have the \emph{weak spatiotemporal chaos}.
We shall denote with a star the values of the parameters for which the
synchronization transition occurs, distinguishing the weak and strong cases by
the index $w$ or $s$, respectively.  

From the results of our numerical simulations,  we think that the following
Ansatz for the topological distance $\rho(p)$ holds: the strong chaos
transition always belong to the directed percolation universality class, while
the weak chaos transition is always of first order character. This statement is
equivalent to the assumption that the fraction of sites in the percolation
cluster that  do not belong to the difference cluster is either vanishing
(strong chaos)  or order one (weak chaos) in the thermodynamic limit. If 
noise is allowed to desynchronize the system, the directed percolation
character is lost. 

This Ansatz can be illustrated by plotting the ratio between the topological
distance $\rho(p)$ and the density of wet sites $m(p)$ as shown in
Fig.~\ref{fig:Ansatz} for the logistic map and $\eps=1/3$.  One can see that
this ratio maintains almost constant up to the transition point.  We have
checked that this behavior hold for several values of the slope $a$ and the
coupling $\eps$ for the three maps.

It is interesting to examine the behavior of the difference $z_i^t$ near the
transition point. In the strong chaos phase, the value of the difference
$z_i^t$ for non-synchronized sites is large, since the synchronization if
forced by the percolation mechanism. 
On the other hand, in the weak chaos phase, the value of the difference
$z_i^t$ for non-synchronized sites is small in average, even though localized
bubbling of large difference can be observed~\cite{bubbling}.

In the following we study numerically the phase boundary between the strong and
weak chaos for the Bernoulli shift and the logistic and tent map, and we obtain
some analytical approximation for it. 
 
\section{Phase diagrams}
\label{sec:phasediagram}

The boundary between the weakly and strongly chaotic phases 
can be defined considering the synchronization mechanism
on the critical percolation cluster, i.e.\ the cluster of wet sites at
$p^*=\min p^*_s=\max p^*_w=1-p_c^{\mbox{\scriptsize (DP)}}$.
As illustrated in Appendix~A,
the critical wet cluster is included in all percolating wet clusters, so 
it can be considered to be the smallest percolating cluster of wet sites, and
thus the one most unfavorable to the proagation of the difference, still
presenting spanning paths over which chaos can propagate.  
The algorithm for the generation of the critical cluster is also described in 
Appendix~A. 

\begin{figure}
	\centerline{\psfig{figure=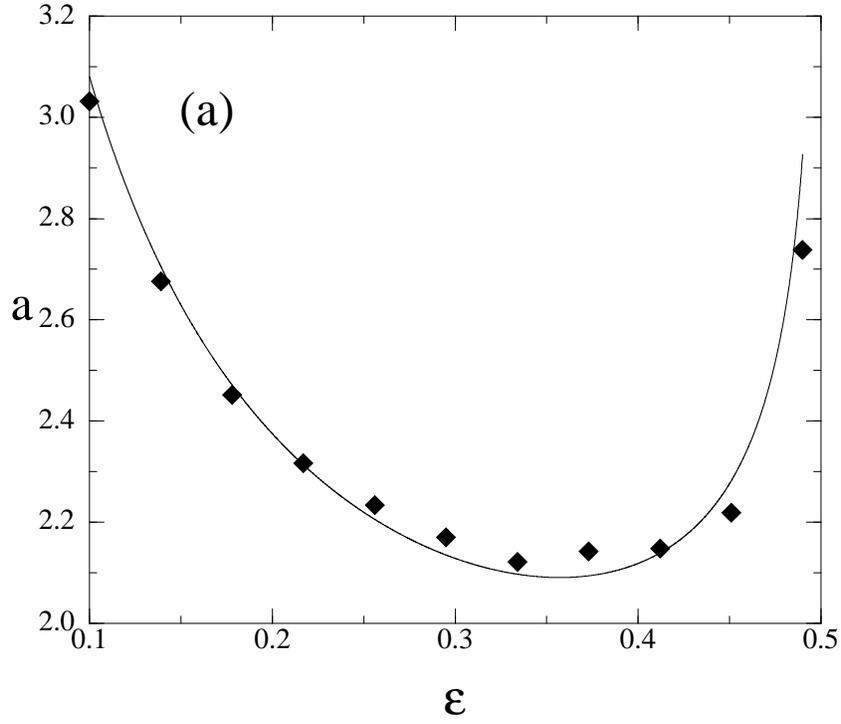,width=11cm,angle=270}}
	
	\centerline{\psfig{figure=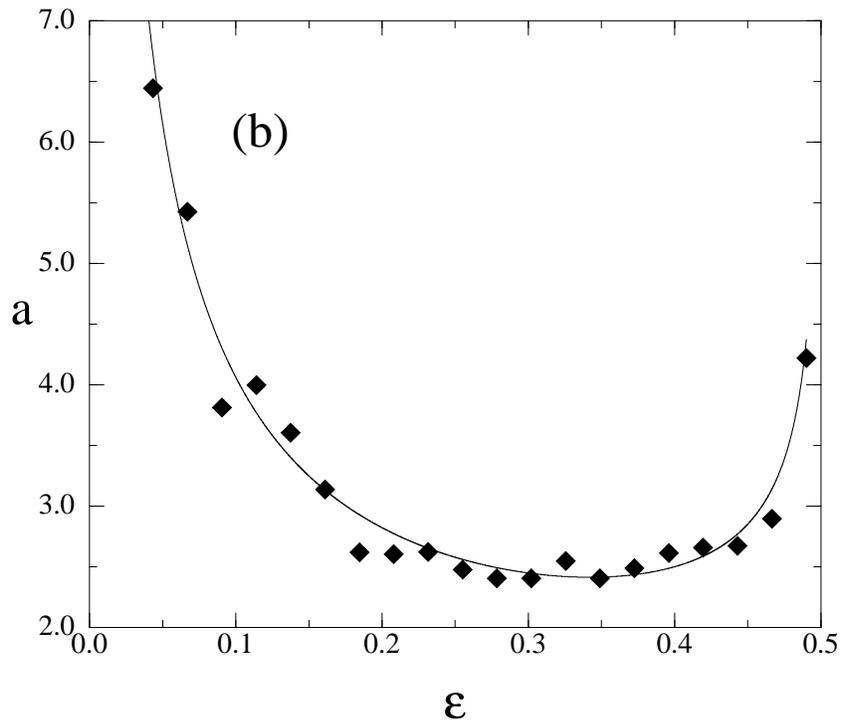,width=11cm,angle=270}}
\caption{Percolative phase diagram for Bernoulli (a) and tent
(b) maps. The line represent a power law fitting, as described in
the text. Average over 5 runs, $N=100$, $T=4000$.}
\label{fig:phasediagram}
\end{figure}

\begin{figure}
	\centerline{\psfig{figure=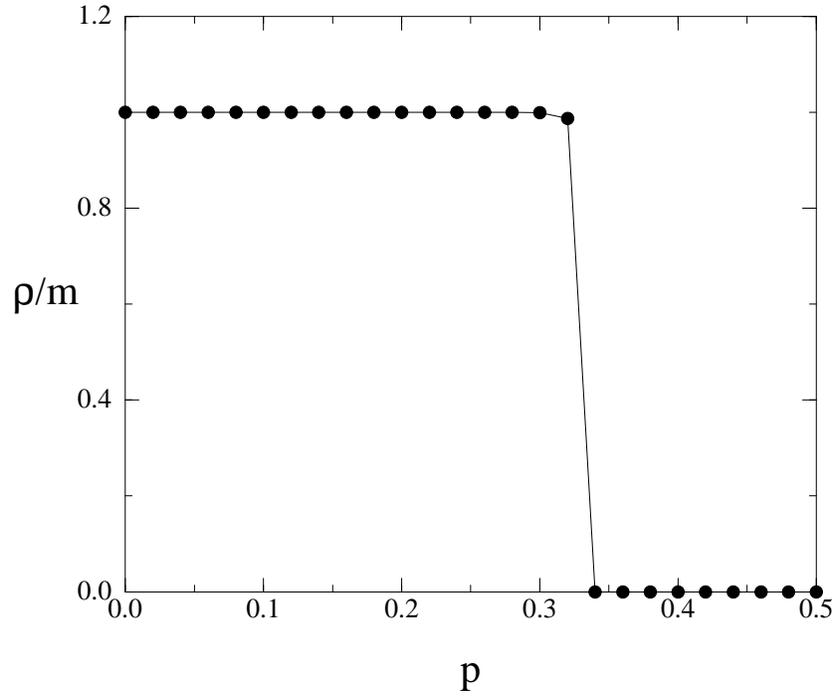,width=11cm,angle=270}}
	\caption{The ratio $\rho(p)/m(p)$ vs. $p$ for a chain of 1000 logistic
	maps, $T=3000$, slope $a=4.$ and $\eps=1/3$.}
	\label{fig:Ansatz}
\end{figure}

The phase diagrams for the Bernoulli shift and the tent map are shown in
Fig.~\ref{fig:phasediagram}; the logistic map never exhibits the strongly
chaotic phase. The case $\eps=1/2$ is special, since the connectivity changes,
and in this case the lattice corresponds to that of the Domany-Kinzel
model~\cite{DK},  for which $1-p_c^{\mbox{\scriptsize (DP)}}=0.3$. 

Since this boundary phase is defined on a critical percolation
cluster, one can assume that the average expansion rate, given by the
sum of all paths in the wet cluster, is dominated by a single path.
Let us assume that this path contains $x$ vertical steps (of weight
$1-2\eps$) and $y$ oblique steps (of weight $\eps$). For the Bernoulli
shift, all steps carry an expansion rate $a$ and thus  one obtains for
the Lyapunov multiplier $\mu_c$ at the phase boundary 
\[
	\mu_c \simeq a^{x+y} (1-2\eps)^x \eps^y  \simeq 1
\] 
and thus 
\[	
	a^* = \dfrac{\mbox{const}}{(1-2\eps)^{\alpha_1} \eps^{\alpha_2}},
\]
with
\begin{equation}
	\alpha_1+\alpha_2=1.
	\label{norm}
\end{equation}
Indeed, the numerical data presented in Fig.~\ref{fig:phasediagram}
does support a correspondence of this form. However, they are too rough to
allow the precise computation of the exponents, which anyhow do not seem to
always obey to Eq.~(\ref{norm}). 

We computed numerically the transition point $p^*$ and the   critical exponents
$\beta$ and $\nu_{\bot}$ for the  topological distance $\rho(p)$ for the
Bernoulli shift and the tent map by means of the scaling relation
\begin{equation}
	\rho\left(p,\dfrac{1}{t}\right)=\alpha^{\frac{\beta}{\nu_{\bot}}}\cdot
	\rho\left(\alpha^{\frac{1}{\nu_{\bot}}}(p-p^*) + p^*,
	\frac{1}{\alpha t}\right)
	\label{invariant}
\end{equation}
that holds in the thermodynamic limit $N\rightarrow\infty$. Here  $\alpha$ is
an arbitrary time scaling factor. We found $p_c = 0.460(2)$, $\beta = 0.26(1)$
and  $\nu_{\bot}=1.75(5)$.  The results are consistent with the hypothesis
that  the phenomenon belongs to the directed percolation universality class, as
expected.

The strong chaos transition is characterized by the expanding dynamics on the
percolation cluster. Thus, in general, at the  transition point the value of a
non-zero difference  $z^t_i$ is not small. On the contrary, the weak chaos
transition is characterized by the average vanishing of $z^t_i$. Up to the
boundary between the two regimes the distance field $z_i$ is still small and a
linear approximation  can be used.  The synchronization mechanism is related to
the average number of links between wet sites, $\kappa(p)$. Since the wetting
is forcing (once that a site has been wet, it cannot be ``unwet''), one can
simply consider the process for a single site, obtaining $\kappa(p) = 3 (1-p)$.

By imposing that  the distance should be marginally expanding, and assuming
that the difference field $z_i$ is almost constant, we get  for the Bernoulli
shift and $\eps=1/3$, 
\[
	\dfrac{1}{3} a^*_w \kappa(p^*_w) = a^*_w (1-p^*_w) = 1
\]
and thus $a^*_w = 1/(1-p^*_w)$. 

For a generic map, with positive and negative slopes, cancellation effects  can
be present. We need an indicator about the constancy of the sign of the
derivative over the interaction range of the diffusion operator. We consider
the quantity
\[
\eta_i = \dfrac{f'(x_i) + f'(x_{i+1})}{2},
\]
which is analogous to the local Lyapunov multiplier, and
\[
\eta = \left(\prod_{i=0}^{N-1} \eta_i\right)^{1/N}
\]
as a global indicator. 
For the Bernoulli map this indicator corresponds to the slope
$a$. 

Thus, for $\eps = 1/3$, the condition for observing the
weak chaos synchronization transition is given in this approximation
by  
\[
\dfrac{1}{3} \cdot \eta^*_w \cdot \kappa(p^*_w) = \eta^*_w (1-p^*_w)=1.
\]

\begin{figure}
		\centerline{\psfig{figure=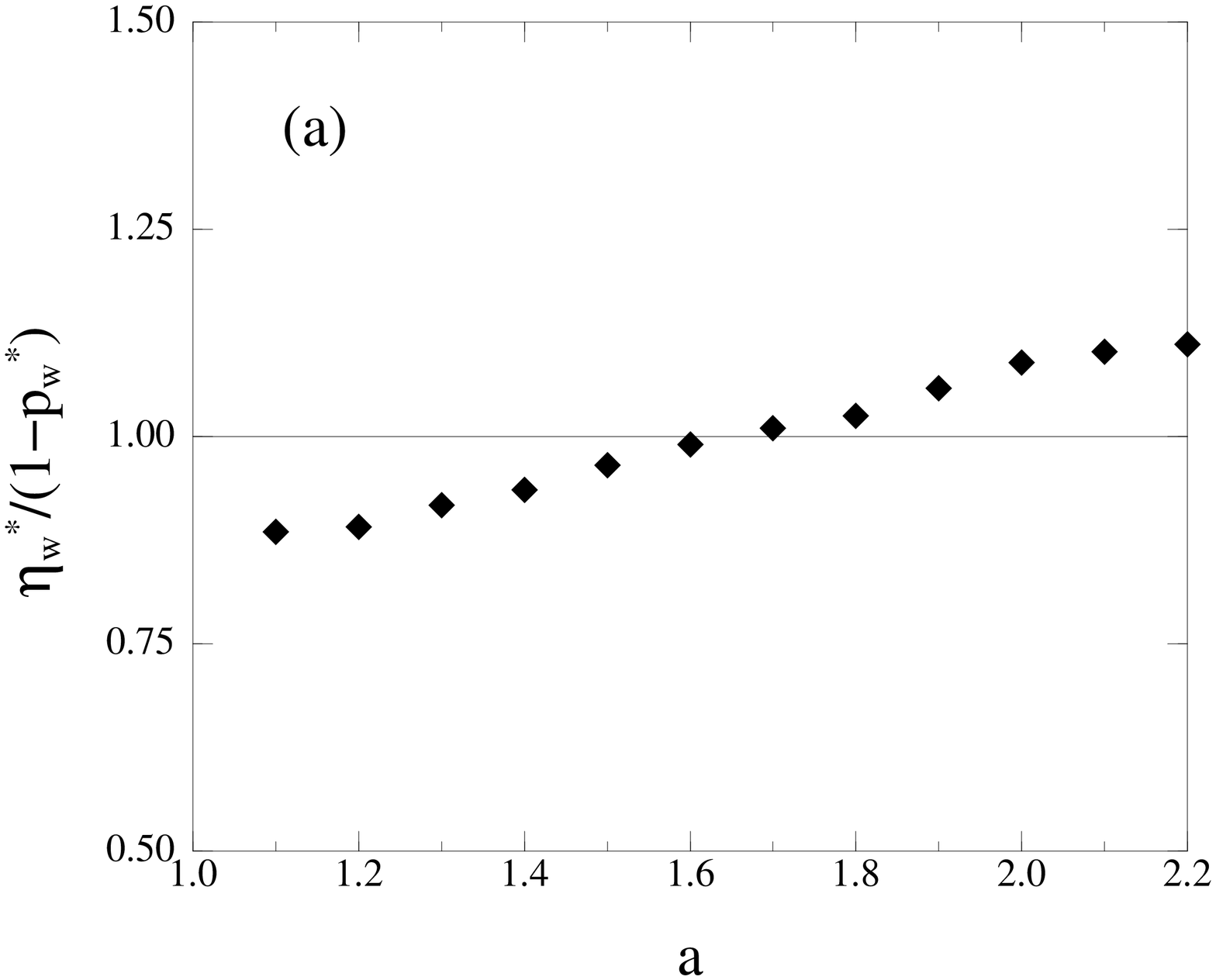,angle=270,width=11cm}}
	
		\centerline{\psfig{figure=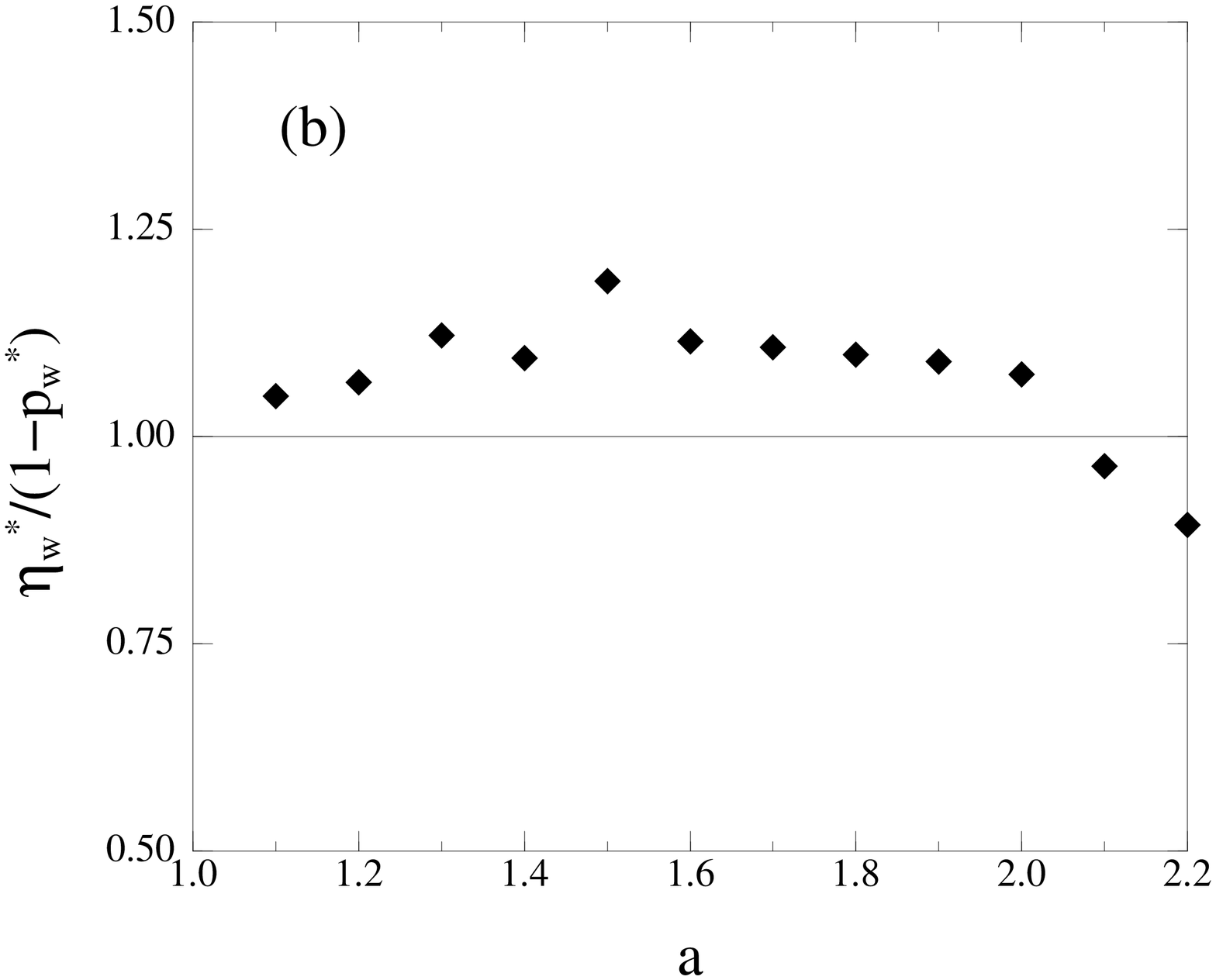,angle=270,width=11cm}}
	
	\caption{\label{fig:weak}
		The relation between the slope $a$ and the weak chaos synchronization 
		threshold $\eta_w^*/(1-p^*_w)$ for a chain of Bernoulli shift
		(a)
		and tent maps (b). Average over 4 runs,
		$N=400$, $T=8000$, $\eps=1/3$.
	}
\end{figure}

We report in Fig.~\ref{fig:weak} the behavior of $\eta^*_w(1-p^*_w)$ as a
function of the slope $a$ for Bernoulli shift and tent maps, together with  the
mean field approximation.  The discrepancies from this approximation originate
from  the assumption  of  a uniformly vanishing difference field $z_i$, which
is not fulfilled even at the transition point.

Let us turn now to the metric distance $\zeta$. 
For random maps even an infinitesimally small distance is amplified
in one step to a random value. Thus, we have for the metric distance
\[
	\zeta (t) = \langle |z|\rangle \rho(t),
\]
and $\zeta$ has the same critical behavior of $\rho$, denoting a
certain degree of universality in this synchronization transition. 

\begin{figure}
		\centerline{\psfig{figure=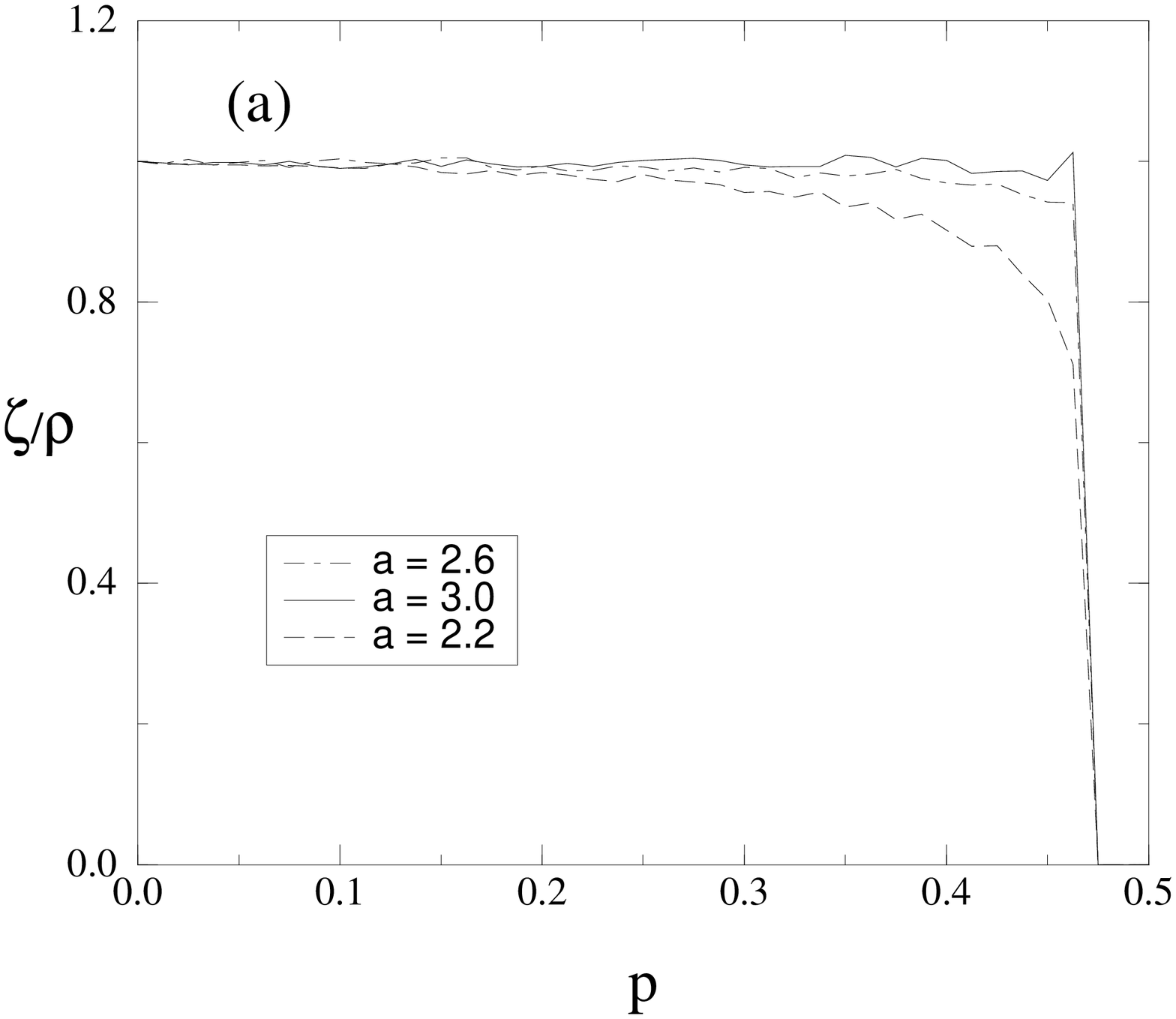,angle=270,width=11cm}}

		\centerline{\psfig{figure=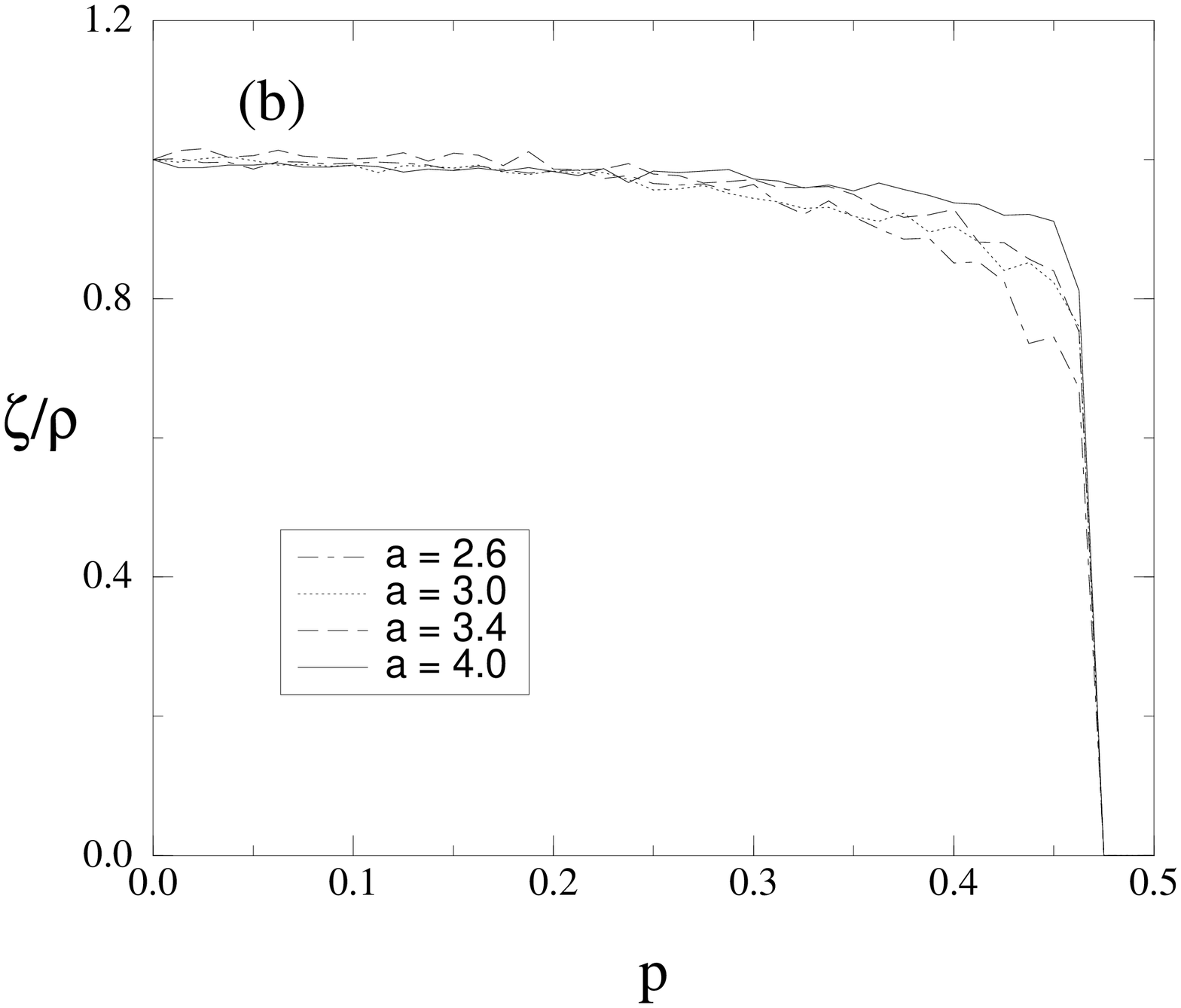,angle=270,width=11cm}}

	\caption{\label{fig:zeta}
		The dependence of the ratio between the metric distance
		$\zeta(p)$ and the topological distance $\rho(p)$ as a
		function of $p$, for the Bernoulli shift (a) and tent map
		(b) and different slopes $a$. Average over 4 runs,
		 $N=100$, $T=1000$.
	}
\end{figure}

We have studied the behavior of the ratio $\zeta(p)/\rho(p)$  as a function of
$p$ for various values of the slope $a$  for the Bernoulli shift and the tent
map, and $\eps=1/3$. The results, reported in Fig.~\ref{fig:zeta}, show that
for the Bernoulli shift the  metric distance is independent of $p$ and of slope
$a$ far from the synchronization transition. Deviations from this behavior near
the transition point   vanish as the slope $a$ become greater than $a^*$.

For tent map the ratio $\zeta(p)/\rho(p)$ is independent of $p$ far for the
transition but it depends on the slope $a$ which determines the distribution of
$z_i$. Again, for $a\gg a^*$, the metric distance is almost constant from $p$
also near the transition point.

\begin{figure}
	\centerline{
		\psfig{figure=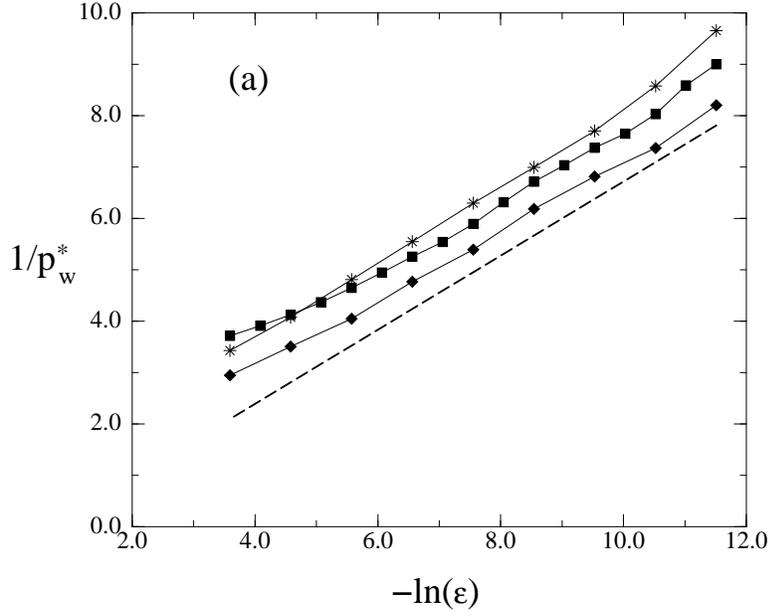,angle=270,width=11cm}
	}
	
	\centerline{	\psfig{figure=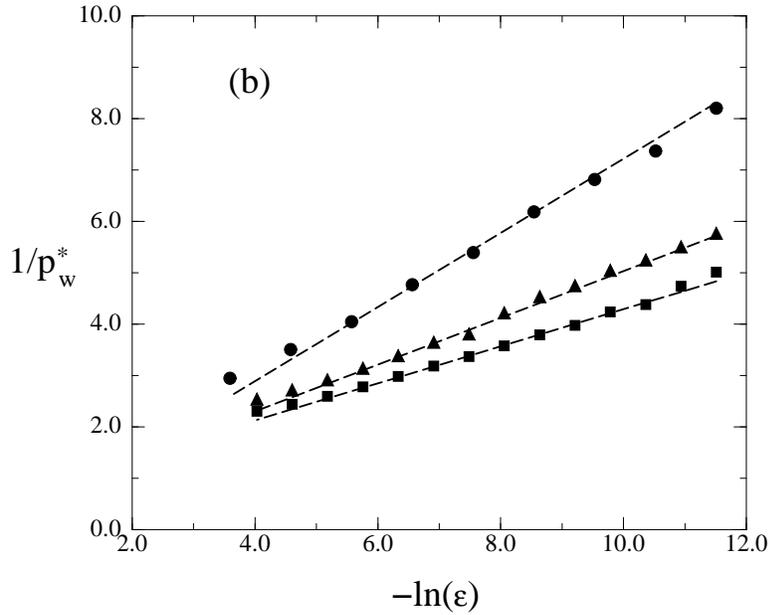,angle=270,width=11cm}
	}
	\caption{\label{fig:piceps}
		Relation between a small coupling $\eps$ and the synchronization
		threshold $p^*_w$. Part (a): the three sets 
		correspond to the Bernoulli shift with slope $a=2$ (stars),
		tent map with slope $a=2$ (diamonds) and logistic maps with
		$a=4$ (squares). For all three maps the maximum 
		Lyapunov exponent for $\eps=0$ is $\lambda=\ln(2)$; the dashed line
		correspods to the law $1/p^*_w = \ln(\eps)/(2\ln(2))$.
		Part (b): Bernoulli shift for $a=2$ (circle), $a=3$
		(triangles) and $a=4$ (squares); the dashed lines have slope
		$1/(2\ln(a))$.	Data from one simulation with $N=500$ and
		$T=2000$.	 	
	}
\end{figure}

The case $\eps\ne 1/3$ is more difficult to analyze, since the asymmetric
couplings cannot be easily mapped onto a statistical problem.   However, for
very small $\eps$, the expansion is dominated by the exponential growth (with
average rate $\lambda$) along the vertical link  of the largest difference. A
very rough mean field description could be the following. Let us assume  that
at a certain time there is essentially only one site $i$ with a non null
difference $z_i$. This difference grows exponentially at rate $\lambda$, and
propagates to the neighboring sites at rate $\eps$. After an average time $1/p$
the difference at site $i$ is set to zero by the synchronization mechanism, and
after time $2/p$ only one of the neighboring has a nonzero difference, thus
closing the cycle.  Thus, the average expansion rate at the synchronization
threshold $p_c$ is  
\begin{equation}
	 \eps \exp(2\lambda/p_c) \simeq \mbox{const},
	 \label{piceps}
\end{equation}
where $\lambda$ is the Lyapunov exponent of the uncoupled map, and we have
neglected non-exponential prefactors.  In Fig.~\ref{fig:piceps} we show the
results of one simulation for the three maps studied in the article, with the
parameters chosen so to have $\lambda=\ln(2)$, and for Bernoulli maps with
various slopes. One can see that Eq.~(\ref{piceps}) is verified for small
$\eps$, except finite size and time effects.  

\section{Conclusions}
We have studied the synchronization transition between two chains of
diffusively coupled chaotic maps, induced by the inactivation of degrees of
freedom in the difference space with a probability $p$. We have found that two
different regimes can be defined: the strong chaos regime for which the
dynamics of the transition is dominated by the directed percolation transition,
and a weak chaos regime in which the system synchronizes in presence of
spanning paths along which the difference could in principle survive. We have
been able to present some analytical approximation of the transition point and
its critical properties.

The character of the transition and the critical value of the parameter $p$ are
proposed as indicators of spatial propagation of chaoticity, which can
complement the usual Lyapunov description. These indicators does not rely on
the existence of a tangent space or exponential growth, so they can be applied
to a broader class of systems, like non differentiable maps or cellular
automata~\cite{BagnoliRechtman}, and system presenting stable
chaos\~cite{stablechaos}.

Our approach can be considered as the annealed version of models that exhibit a
synchronization transition, presented in some recently appeared papers. First
of all, let us consider the synchronization mechanism proposed by Pecora and
Carroll~\cite{PecoraCarroll}. In their numerical and  experimental set-up they
studied the behavior of the distance between two chaotic oscillators, when part
of the degrees of freedom of one of them is set equal to the corresponding
degrees of freedom of the other. Extending this mechanism to spatially extended
systems, one has a coupling similar to ours, but with quenched disorder (the
coupling degrees of freedom). We did not perform the study of quenched disorder
in diffusively coupled maps, since it implies longer spatial couplings in order
to avoid the formation of walls.

Another similar system was studied by Fahy and Hammann~\cite{Fahy}. Their
subject was an ensemble of non-interacting particles in a chaotic potential. At
fixed time intervals the velocities of the particles were all set equals to a
Gaussian sample. If the free-fly time was small enough, all trajectories
collapse into one. We performed preliminaries simulations (not reported here)
on a modified system, in which one reference particle followed an unperturbed
trajectory, while a replica had its velocity set equal to that of the reference
at time intervals $\tau$. Indeed, we observed a synchronization transition for
small enough $\tau$. By observing the system stroboscopically at intervals
$\tau$ we can substitute the continuous dynamics with a map. In this case the
time $\tau$ controls the chaoticity of the map, while the synchronization
mechanism is similar to that of Pecora and Carroll and thus to a quenched
disorder. Fahy and Hammann also checked that the synchronization transition
occurs if the position of the particles is set equal, instead of the velocity.

The synchronization mechanism studied in this article is quite particular,
since it implies a complete collapse of the distance between the master and the
slave.  The strong chaos synchronized phase is not stable with respect to the
inclusion of desynchronizing effects, such as noise or non perfect identity of
parameters in the master and slave systems. On the other hand, the weak chaos
transition is ruled by the exponential shrinking of difference, i.e. by
negative Lyapunov exponents for the difference. Thus, it is expected that this
transition is robust with respect to weak desynchronization effects, and not to
depend on the threshold $\tau$. The modification of the observed phenomena in
presence of noise will be the subject of a future work. In the present
version, the synchronization transition can be considered as a
mathematical tool for the definition of quantities related to the spatial
propagation of chaoticity. The most natural systems to which this method can be
applied are those presenting stable chaos~\cite{stablechaos}, i.e.\ irregular
behavior in the presence of negative Lyapunov spectra. For these systems an
eventual small noise if wiped out by the contracting dynamics on small scales.
Similar noise-free systems are those that can be approximated by cellular
automata models~\cite{BagnoliRechtman}.

\section*{acknowledgements}
We thank S. Ruffo, R. Rechtman and M. Bezzi for fruitful discussions.
Part of the  simulations have been performed on the CRAY-T3E of the CINECA
center using the INFM facilities ({\it Progetto Calcolo Parallelo}). L.B. and
P.P. thank the {\it Dipartimento di Matematica Applicata} for kind hospitality.
Part of this work was performed during the workshop ``Complexity and Chaos'' at
ISI Foundation, Torino, Italy.

\section*{Appendix~A}
The generation of the critical cluster can be done in a very efficient way,
using a modification of the invasion percolation algorithm~\cite{invasion}.

The random bit $r_i^t(p)$ is obtained by comparing a random number $R_i^t$,
whose distribution probability is constant in the unit interval, with $p$. If
$R_i^t<p$ then $r_i^t=1$, else $r_i^t=0$.

Let us consider a lattice with the same geometry of the percolation one and
assign to each site a random number $R_i^t$. The idea now is that of lowering
$p$ (starting from $p=1$) until the cluster of $r_i^t=0$ sites spans the
lattice. A nice property of the directed site percolation problem (which is
related to the ``forcing'' character of wetting) is that, given the random
number $R_i^t$, the percolation cluster for $p_1$ is included in the
percolation cluster for $p_2$ if $p_1<p_2$. Thus, one has simply to choose the
site with the highest $R_i^t$ on the border of percolation cluster (i.e.\ among
the sites with $r_i^t=1$ connected to some $r_j^{t-1}=0$ site, with $|i-j|=1$).
This maximum value will become the new estimate of $p$, and the percolation
cluster is enhanced to include all sites with $R_i^t>p$.

This procedure can be easily performed by keeping the values of the sites in
the border in an ordered linked list, assuming that all sites in the $t=0$ row
are connected to a wet site.

\newpage

\vfill

F. Bagnoli, L. Baroni and P. Palmerini, {\it Synchronization and DP
in CML} \hfill Figure 1

\newpage 

\vfill

F. Bagnoli, L. Baroni and P. Palmerini, {\it Synchronization and DP
in CML} \hfill Figure 2

\newpage

\vfill

F. Bagnoli, L. Baroni and P. Palmerini, {\it Synchronization and DP
in CML} \hfill Figure 3

\newpage

\vfill

F. Bagnoli, L. Baroni and P. Palmerini, {\it Synchronization and DP
in CML} \hfill Figure 4

\newpage

\vfill

F. Bagnoli, L. Baroni and P. Palmerini, {\it Synchronization and DP
in CML} \hfill Figure 5

\newpage

\vfill

F. Bagnoli, L. Baroni and P. Palmerini, {\it Synchronization and DP
in CML} \hfill Figure 6

\newpage

\vfill

F. Bagnoli, L. Baroni and P. Palmerini, {\it Synchronization and DP
in CML} \hfill Figure 7

\end{document}